# Sequences of selective rotation operators to engineer interactions for quantum annealing on three qutrits

V.E. Zobov[a*], I.S. Pichkovskiy[b]

[a]Kirensky Institute of Physics, Federal Research Center KSC SB RAS, Krasnoyarsk, 660036 Russia;
[b]Institute of Engineering Physics and Radio Electronics, Siberian Federal University, Krasnoyarsk, 660074 Russia

**ABSTRACT**

We have done simulating of factorization the number 15 on three qutrits, represented by the spins $S = 1$, by quantum annealing. We assume that strong one-spin interaction allow selectively operate on different transitions between levels of the each qutrit. We present a sequence of selective rotation operators to engineer from dipole-dipole interaction a time-dependent effective Hamiltonian necessary for solving the problem. Also we find dependence of fidelity versus different parameters: magnetic field, total annealing time, and duration of time step, when the continuous variation of the Hamiltonian is replaced by a discrete one.

## 1. INTRODUCTION

At present, the efforts of many researchers are aimed at developing a quantum computer[1,2], which promises a significant acceleration of computations compared to a conventional classical computer. Quantum calculations on it can be done in two ways: firstly, using a network of elementary logical operators[1] (gates), secondly, by means of a slow (adiabatic) change in time of the Hamiltonian from the initial form, the ground state of which is easy to prepare, to the final form, in the ground state of which the solution of the problem is encoded[3,4]. According to the theory, both models are equally effective in solving complex problems, but it is believed that adiabatic quantum calculations are more resistant to some perturbations.

An important variant for the practice of adiabatic quantum computing is quantum annealing[5,6], in which the Hamiltonian of the Ising model in a transverse magnetic field is taken as the time-dependent Hamiltonian. At the initial moment of time, the magnitude of the interaction with the field is many times greater than the magnitude of the spin-spin interaction, providing mixing of all possible states of the latter. At the final moment of time, the field is turned off, and the spin system is in the state with the minimum energy of the Ising model, the interaction constants of which are prepared in accordance with the problem to be solved.

As elements, carriers of quantum information, as a rule, consider two-level quantum systems – qubits[1]. The operations performed on them are described in binary number system. The same operations can be performed by taking as the elements quantum systems with three levels – qutrits[7-11]. Qutrits promise an increase in the efficiency of quantum computing: firstly, due to a faster increase in the size of the computational basis (Hilbert space) with an increase in the number of elements, secondly, through the use of a ternary number system, which is considered more efficient than binary[9-12].

As qutrits, it is proposed to use, for example, objects with spin $S = 1$ in magnetic and crystal fields. These include quadrupole nuclei[7,13] of deuterium, nitrogen or lithium, as well as NV centers in diamond (paramagnetic color centers formed by an electron on a vacancy near a nitrogen atom)[14]. The latter variant is preferable because of the presence of a strong dipole-dipole interaction (DDI) between NV centers, which is necessary for the implementation of conditional operations in quantum algorithms. Moreover, in this case, there is a large difference in the frequency of transitions between different energy levels, which makes it possible to control the states of the system with the help of transition-selective microwave (MW) field pulses.

---

[*] rsa@iph.krasn.ru

The development of control methods for many-body quantum systems is the most important direction in the implementation of a quantum computer. To date, variants of sequences of selective rotation operators for performing gates on a separate qutrit[7,8] and two qutrits[8,13,15] have been proposed. In the work[10], sequences for the adder in the ternary number system on the chain of qutrits were obtained. In the work[14], for an ensemble of NV centers in diamond that are connected by a DDI, sequences of transitions-selective MW pulses were obtained, which can be used to eliminate the DDI or convert it into two-spin interactions of another type. The adiabatic quantum calculation was not considered in these papers. The simulation of the adiabatic quantum factorization algorithm on two qudits (*d*-level quantum systems) was performed in the work[16], and sequences of selective RF pulses needed to create a time-varying effective Hamiltonian were found. In the present work, we simulate the solution of the factorization problem by quantum annealing on three qutrits. We found sequences of rotation operators, selective for transitions and spins, which allow to switch off unnecessary DDI bonds and create effective interactions of the desired type, including three-spin interactions. Note that, for $S = 1/2$, a method for creating a three-spin interaction was proposed[17]. This approach, based on the properties of the Pauli matrices, is not applicable for $S = 1$. We are based on the results of the theory of coherent averaging in the NMR of a solid[18], in which the RF field pulses eliminate the DDI in the first approximation. Therefore, the spin dynamics is determined by multi-spin interactions obtained in higher orders of the theory of the average Hamiltonian.

## 2. SYSTEM OF THREE SPINS $S = 1$

We will consider a system of three qutrits represented by spins with $S = 1$ with different resonant frequencies due to magnetic and crystal fields and connected by a dipole-dipole interaction (DDI). Hamiltonian is

$$H = H_1 + H_d, \tag{1}$$

$$H_1 = -\omega_1 S_1^z - \omega_2 S_2^z - \omega_3 S_3^z + q_1\left[(S_1^z)^2 - \frac{2}{3}\right] + q_2\left[(S_2^z)^2 - \frac{2}{3}\right] + q_3\left[(S_1^z)^2 - \frac{2}{3}\right], \tag{2}$$

$$H_d = J_{12} S_1^z S_2^z + J_{13} S_1^z S_3^z + J_{23} S_2^z S_3^z, \tag{3}$$

where $S_j^z$ is the operator of the projection of the spin *j* on the Z axis with the eigenvalues $m_j$ taking one of the three values 1, 0, -1. We denote the corresponding eigenfunctions $|m_j\rangle$. In this basis, single-spin energy takes three values respectively (three energy levels): $\varepsilon_{j1} = -\omega_j + q_j/3$, $\varepsilon_{j2} = -q_j/3$, $\varepsilon_{j3} = \omega_j + q_j/3$. The state of the system can be controlled using microwave pulses with frequencies equal to the differences in the energy of two levels $\omega_{MW} = \omega_{jnk} = \varepsilon_{jn} - \varepsilon_{jk}$, which we will call transition frequencies. *k* and *n* are numbers of selected levels, *j* is number of spin. For successful control, we assume that the difference in the magnitudes of the resonant frequencies of different transitions in all three spins exceeds many times the magnitude of the DDI. In this case, the MW pulse will act only on the selected resonant transition, causing coherent changes in the corresponding two states. These changes are described by operators that coincide with rotation operators of a two-level system (with effective spin $S = 1/2$), which are called selective rotation operators[14,16,19] $\{\Omega\}_{\alpha,j}^{k\leftrightarrow n}$ and in the matrix representation have the form:

$$\{\Omega\}_{z,j}^{1\leftrightarrow 2} = \begin{pmatrix} \exp\left[-i\frac{\Omega}{2}\right] & 0 & 0 \\ 0 & \exp\left[i\frac{\Omega}{2}\right] & 0 \\ 0 & 0 & 1 \end{pmatrix}, \quad \{\Omega\}_{z,j}^{2\leftrightarrow 3} = \begin{pmatrix} 1 & 0 & 0 \\ 0 & \exp\left[-i\frac{\Omega}{2}\right] & 0 \\ 0 & 0 & \exp\left[i\frac{\Omega}{2}\right] \end{pmatrix},$$

$$\tag{4}$$

$$\{\Omega\}_{y,j}^{1\leftrightarrow 2} = \begin{pmatrix} \cos\frac{\Omega}{2} & -\sin\frac{\Omega}{2} & 0 \\ \sin\frac{\Omega}{2} & \cos\frac{\Omega}{2} & 0 \\ 0 & 0 & 1 \end{pmatrix}, \quad \{\Omega\}_{y,j}^{2\leftrightarrow 3} = \begin{pmatrix} 1 & 0 & 0 \\ 0 & \cos\frac{\Omega}{2} & -\sin\frac{\Omega}{2} \\ 0 & \sin\frac{\Omega}{2} & \cos\frac{\Omega}{2} \end{pmatrix},$$

where $\Omega$ is the angle of rotation around the axis $\alpha$ ($\alpha = x, y, z$). To implement the selective rotation on the angle $\Omega = h_f t_p$ we can switch on a time $t_p$ the MW field with amplitude $h_f$ (in frequency units) and frequency $\omega_{MW} = \omega_{jnk}$ ($t_p \gg 1/\omega_{mw}$). The X-rotation matrix differs from the Y-rotation by the coefficients (-i) before both sines. Last, we can make a non-selective spin rotation[14,19], acting during the time $t_p$ simultaneously on two transitions by two MW fields with frequencies $\omega_{12}, \omega_{23}$ and amplitude $h_f$.

## 3. ADIABATIC ALGORITHM FOR FACTORIZATION WITH THREE QUTRITS.

Let us consider the factorization of the number 15 by means of quantum annealing[6], otherwise through evolution of system with Hamiltonian[6,20,21]:

$$H(t) = \left(1 - \frac{t}{T}\right)H_0 + \left(\frac{t}{T}\right)H_p, \qquad 0 \leq t \leq T. \qquad (5)$$

where $H_P = (n - pq)^2$ is target Hamiltonian, in which numbers $p$ and $q$ should be expressed through spin variables of our system. The ground state for which $p = 5$ and $q = 3$, has zero energy. Whereas for other states, the energy will assume large positive values. Following of work[20], we chose multipliers among the odd numbers: $p = 2a + 1$, $q = 2b + 1$, but the numbers $a$ and $b$ were presented not in binary but in symmetric ternary number system: $a = 3a_1 + a_2$, $b = b_0$. For recording numbers we use a computational basis of the eigenfunctions $|m_1, m_2, m_3\rangle$ of the spin projection operators on the axis Z: $S_1^z, S_2^z, S_3^z$. Each of the projections may have meaning 1, 0, -1. In this basis:

$$a_1 = S_1^z, \; a_2 = S_2^z, \; b = b_0 = S_3^z. \qquad (6)$$

After transformations we get:

$$H_P = (15 - (6a_1 + 2a_2 + 1)(2b + 1))^2 = 144b^2 a_1^2 + 96b^2 a_1 a_2 + 16b^2 a_2^2 + 48b^2 a_1 + 16b^2 a_2 + 144ba_1^2 + \\ + 96ba_1 a_2 + 16ba_2^2 + 4b^2 - 312ba_1 - 104ba_2 + 36a_1^2 + 24a_1 a_2 + 4a_2^2 - 56b - 168a_1 - 56a_2 + 196 \qquad (7)$$

As an initial interaction, we took the interaction of spins with a transverse magnetic field directed along the axis X:

$$H_0 = -h\left(S_1^x + S_2^x + S_3^x\right). \qquad (8)$$

As an initial state we use ground state $|\psi\rangle$ of this Hamiltonian, which is direct product of eigenvectors with eigenvalue 1 of separate spins: $S_1^x, S_2^x, S_3^x$.

The solution of our problem $|\Psi\rangle$ will be founded in the following form:

$$|\Psi\rangle = \hat{Q}\exp\left(-i\int_0^T H(t)dt\right)|\psi\rangle \cong \prod_{l=0}^{N} U_l |\psi\rangle, \qquad (9)$$

where $\hat{Q}$ is time-ordering operator. Following the works[16,20,21], we represented the operator of adiabatic evolution during time $T = \Delta t N$ with linear Hamiltonian (5) as a product of evolution operators on the sequence of N small time intervals $\Delta t$. On each such interval, we neglect the variation of Hamiltonian (5) and approximately represent the evolution operators as product of two non-commuting operators:

$$U_l = \exp\left[-i\left(1-\frac{l}{N}\right)\Delta t H_0\right]\exp\left[-i\Delta t H_p \frac{l}{N}\right], \tag{10}$$

where *l* is discreet time (0≤*l*≤*N*).

## 4. ENGINEERING OF EFFECTIVE INTERACTION

The first factor $\exp\left[-i\left(1-\frac{l}{N}\right)\Delta t H_0\right]$ in evolution operators $U_l$ (10) we will received as a non-selective rotation of the spin[14,19], given by the operator $\exp(-i\theta S^x)$, by an angle $\theta = h\Delta t\left(1-\frac{l}{N}\right) = t_p h_f$. Consider the creation of the second factor $\exp(-i\Delta t H_p\, l/N)$ in the evolution operator (10). We divide it into the product of evolution operators of each interaction in Hamiltonian $H_p$ (7). This is possible, because all operators in (7) commutate with each other.

### 4.1 Engineering of operators with one spin interaction

Let's start with the terms $-56b - 168a_1 - 56a_2$ in (7). The corresponding evolution operators can be obtained with formula[16]:

$$\exp\left[-i\Omega S_k^z\right] = \{2\Omega\}_{z,k}^{1\leftrightarrow 2}\{2\Omega\}_{z,k}^{2\leftrightarrow 3}, \tag{11}$$

where *k* is number of spin. For example:

$$\exp\left[56i\Delta t\frac{l}{N}a_2\right] = \left\{-112\Delta t\frac{l}{N}\right\}_{z,2}^{1\leftrightarrow 2}\left\{-112\Delta t\frac{l}{N}\right\}_{z,2}^{2\leftrightarrow 3}. \tag{12}$$

For the operators $36a_1^2 + 4a_2^2$ in (7), the corresponding evolution operators can be obtained from[16]:

$$\exp\left[-i3\phi\left(S_k^z\right)^2\right] = \{2\phi\}_{z,k}^{1\leftrightarrow 2}\{-2\phi\}_{z,k}^{2\leftrightarrow 3}\exp[-i2\phi I] \tag{13}$$

$$\begin{pmatrix}\exp[-3i\varphi] & 0 & 0 \\ 0 & 1 & 0 \\ 0 & 0 & \exp[-3i\varphi]\end{pmatrix} =$$

$$\begin{pmatrix}\exp[-i\varphi] & 0 & 0 \\ 0 & \exp[i\varphi] & 0 \\ 0 & 0 & 1\end{pmatrix}\begin{pmatrix}1 & 0 & 0 \\ 0 & \exp[i\varphi] & 0 \\ 0 & 0 & \exp[-i\varphi]\end{pmatrix}\begin{pmatrix}\exp[-2i\phi] & 0 & 0 \\ 0 & \exp[-2i\phi] & 0 \\ 0 & 0 & \exp[-2i\phi]\end{pmatrix}. \tag{14}$$

*I* in (13) is unit matrix. For example:

$$\exp\left[-4i\Delta t\frac{l}{N}(a_2)^2\right] = \left\{\frac{8l}{3N}\Delta t\right\}_{z,2}^{1\leftrightarrow 2}\left\{-\frac{8l}{3N}\Delta t\right\}_{z,2}^{2\leftrightarrow 3}\exp\left[-\frac{8l}{3N}i\Delta t I\right]. \tag{15}$$

### 4.2 Engineering of operators with two-spin interaction

First of all, as in the work[21], in evolution operator with DDI (3) we leave one necessary interaction and destroy two unnecessary interactions using inversion operator:

$$P_k^{-1}S_k^z P_k = \{-\pi\}_{y,k}^{1\leftrightarrow 2}\{-\pi\}_{y,k}^{2\leftrightarrow 3}\{-\pi\}_{y,k}^{1\leftrightarrow 2}S_k^z\{\pi\}_{y,k}^{1\leftrightarrow 2}\{\pi\}_{y,k}^{2\leftrightarrow 3}\{\pi\}_{y,k}^{1\leftrightarrow 2} = -S_k^z, \tag{16}$$

$$\exp(itH_d)P_1^{-1}\exp(itH_d)P_1 = \exp(i2tJ_{23}S_2^z S_3^z), \tag{17}$$

$$P_2^{-1}\exp(itH_d)P_2 P_1^{-1}\exp(itH_d)P_1 = \exp(-i2tJ_{12}S_2^z S_1^z). \tag{18}$$

The following values were accepted for certainty:

$$J_{12} = 24,\ J_{13} = 312,\ J_{23} = 104,\quad H_d = 24a_1a_2 + 104ba_2 + 312ba_1. \tag{19}$$

To simulate (5) - (7), we introduced conditional dimensionless units of measure, which can be easily expressed through the interaction constants of the real system. The necessary numerical value of the exponents of the corresponding factors in $\exp\left[-i\Delta t H_p \frac{l}{N}\right]$ is achieved from (17) and (18) by setting the length of evolution intervals $t$. The length $t$ is chosen[16] to be a multiple of periods $2\pi/\varepsilon_{jk}$.

The operators $48b^2 a_1 + 16b^2 a_2 + 144ba_1^2 + 16ba_2^2$ in $H_p$ can be obtained with formula[16]:

$$\exp\left[-3itJS_p^z(S_q^z)^2\right] = \exp\left[-2itJS_p^z\right]\{-\pi\}_{y,q}^{2\leftrightarrow 3} \exp\left[-itJS_p^z S_q^z\right]\{-\pi\}_{y,q}^{1\leftrightarrow 2} \exp\left[-itJS_p^z S_q^z\right]\{\pi\}_{y,q}^{1\leftrightarrow 2}\{\pi\}_{y,q}^{2\leftrightarrow 3}, \quad (20)$$

where $p$ and $q$ are numbers of spins. For example:

$$\exp\left[-16i\Delta t \frac{l}{N} b(a_2)^2\right] = \left\{\frac{64l}{3N}\Delta t\right\}_{z,3}^{1\leftrightarrow 2} \left\{\frac{64l}{3N}\Delta t\right\}_{z,3}^{2\leftrightarrow 3} \{-\pi\}_{y,2}^{2\leftrightarrow 3} P_1^{-1} \times$$
$$\times \exp\left[-i\frac{l}{N}\frac{\Delta t}{2}\left(\frac{2}{39}\right)H_d\right] P_1 \exp\left[-i\frac{l}{N}\frac{\Delta t}{2}\left(\frac{2}{39}\right)H_d\right] \times \{-\pi\}_{y,2}^{1\leftrightarrow 2} P_1^{-1} \times \quad (21)$$
$$\times \exp\left[-i\frac{l}{N}\frac{\Delta t}{2}\left(\frac{2}{39}\right)H_d\right] P_1 \exp\left[-i\frac{l}{N}\frac{\Delta t}{2}\left(\frac{2}{39}\right)H_d\right] \{\pi\}_{y,2}^{1\leftrightarrow 2}\{\pi\}_{y,2}^{2\leftrightarrow 3}$$

The operators $144b^2 a_1^2 + 16b^2 a_2^2$ in $H_p$ are obtained by a double application of formula (20) (first to one spin, and then to another spin). For example:

$$\exp\left[-16i\Delta t \frac{l}{N}(b)^2(a_2)^2\right] = \left\{\frac{64l}{9N}i\Delta t\right\}_{z,2}^{1\leftrightarrow 2} \left\{-\frac{64l}{9N}i\Delta t\right\}_{z,2}^{2\leftrightarrow 3} \exp\left[-\frac{64l}{9N}i\Delta t I\right] \times$$
$$\times \{-\pi\}_{y,3}^{2\leftrightarrow 3}\left\{-\frac{64l}{9N}i\Delta t\right\}_{z,3}^{1\leftrightarrow 2}\left\{\frac{64l}{9N}i\Delta t\right\}_{z,3}^{2\leftrightarrow 3}\{-\pi\}_{y,2}^{2\leftrightarrow 3}\exp\left[-\frac{16l}{9N}i\Delta t ba_2\right]\{-\pi\}_{y,2}^{1\leftrightarrow 2} \times$$
$$\times \exp\left[-\frac{16l}{9N}i\Delta t ba_2\right]\{\pi\}_{y,2}^{1\leftrightarrow 2}\{\pi\}_{y,2}^{2\leftrightarrow 3}\{-\pi\}_{y,3}^{1\leftrightarrow 2}\left\{-\frac{64l}{9N}i\Delta t\right\}_{z,3}^{1\leftrightarrow 2}\left\{\frac{64l}{9N}i\Delta t\right\}_{z,3}^{2\leftrightarrow 3} \times \quad (22)$$
$$\times \{-\pi\}_{y,2}^{2\leftrightarrow 3}\exp\left[-\frac{16l}{9N}i\Delta t ba_2\right]\{-\pi\}_{y,2}^{1\leftrightarrow 2}\exp\left[-\frac{16l}{9N}i\Delta t ba_2\right]\{\pi\}_{y,2}^{1\leftrightarrow 2}\{\pi\}_{y,2}^{2\leftrightarrow 3}\{\pi\}_{y,3}^{1\leftrightarrow 2}\{\pi\}_{y,3}^{2\leftrightarrow 3}$$

To obtain the final expression from (22), we use formula (17).

### 4.3 Engineering of three-spin interaction in $H_p$

In $H_p$ there are two terms with a three-spin interaction: $96b^2 a_1 a_2 + 96ba_1 a_2$. To create an evolution operator in (10) with a three-spin interaction from the DDI (3), we will follow the ideas of coherent averaging theory[18]

$$\exp\left[ib_{12}S_1^x S_2^z\right]\exp\left[ib_{13}S_1^z S_3^z\right]\exp\left[-ib_{12}S_1^x S_2^z\right]\exp\left[-ib_{13}S_1^z S_3^z\right] \approx \exp\left[ib_{12}b_{13}S_1^y S_2^z S_3^z\right], \quad (23)$$

where $b_{ij} = \Delta t J_{ij} \ll 1$. In (23), terms of the second order of smallness are retained and terms of the third order of smallness are discarded. The required projections of the spin operators are obtained by means of nonselective rotations:

$$\exp\left[ib_{12}b_{13}S_1^z S_2^z S_3^z\right] \approx \exp\left[-i\frac{\pi}{2}S_1^x\right]\exp\left[-i\frac{\pi}{2}S_1^y\right]\exp\left[ib_{12}S_1^z S_2^z\right]\exp\left[i\frac{\pi}{2}S_1^y\right] \times$$
$$\times \exp\left[ib_{13}S_1^z S_3^z\right]\exp\left[-i\frac{\pi}{2}S_1^y\right]\exp\left[-ib_{12}S_1^z S_2^z\right]\exp\left[i\frac{\pi}{2}S_1^y\right]\exp\left[-ib_{13}S_1^z S_3^z\right]\exp\left[i\frac{\pi}{2}S_1^x\right] \quad (24)$$

On the basis of formula (24) we obtain

$$\exp\left[-96i\Delta t \frac{l}{N} b a_1 a_2\right] = \exp\left[i\frac{\pi}{2}S_1^x\right]\exp\left[i\frac{\pi}{2}S_1^y\right]\exp\left[-i\sqrt{96\frac{l}{N}\Delta t}\,a_1 a_2\right]\exp\left[-i\frac{\pi}{2}S_1^y\right]\times$$
$$\times \exp\left[-i\sqrt{96\frac{l}{N}\Delta t}\,a_1 b\right]\exp\left[i\frac{\pi}{2}S_1^y\right]\exp\left[-i\sqrt{96\frac{l}{N}\Delta t}\,a_1 a_2\right]\exp\left[-i\frac{\pi}{2}S_1^y\right]\exp\left[i\sqrt{96\frac{l}{N}\Delta t}\,a_1 b\right]\exp\left[-i\frac{\pi}{2}S_1^x\right] \quad (25)$$

To extract the required pair of spins from the DDI (19), we use formula (17) or (18) in (25).

To engineer the evolution operator $\exp\left[-96i\Delta t \frac{l}{N}(b)^2 a_1 a_2\right]$, we first apply formula (20), and then (24).

To reduce the error, we will divide the time interval into seven additional parts:
$$\exp[-it J_{123} b a_1 a_2] = \{\exp[-i(t/7) J_{123} b a_1 a_2]\}^7. \quad (26)$$

## 5. CALCULATIONS AND DISCUSSIONS

We have found the complete sequence of selective operators of rotations and evolution intervals with the DDI (3), which is necessary for modeling. The result of the calculation is obtained as a superposition of 27 states of the computational basis:
$$|\Psi\rangle = \sum_{m_1,m_2,m_3} C_{m_1,m_2,m_3} |m_1,m_2,m_3\rangle. \quad (27)$$

The accuracy (fidelity) achieved in our calculation will be characterized by the quantity:
$$R = |\langle \Psi | 1,-1,1 \rangle|^2 = |C_{1,-1,1}|^2. \quad (28)$$

The results obtained for different values of the parameters are shown in the figures.

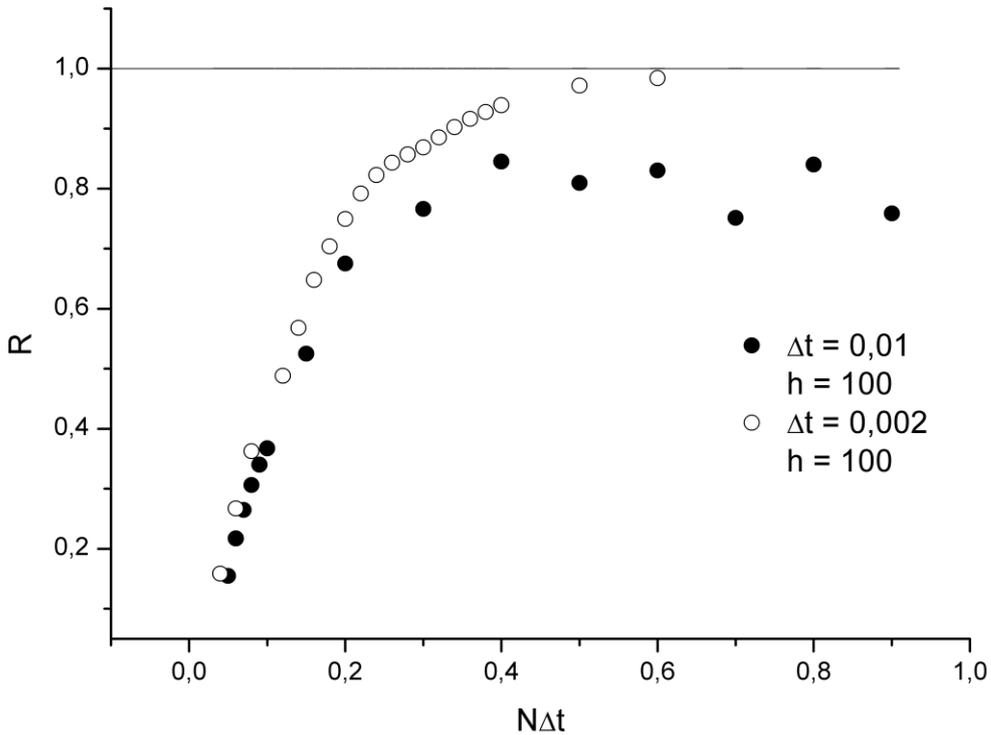

Figure 1 Dependences of the accuracy of factorization $R$ on the annealing time $T=\Delta t N$

On Fig.1 we can see that for small $\Delta t$ and large $N$ we reach an exact solution. This indicates, on the one hand, the correctness of the sequence of operators found, and, on the other hand, the fulfillment of the adiabatic condition. If we used a larger value of $\Delta t$, then the increase in the accuracy of $R$ stops at some limiting residual value due to an error in replacing the continuous change of the Hamiltonian by a discrete in (9) and (10). This residual value depends on the magnitude of the field $h$, as shown in Fig. 2. At low values of the field we can observed increase accuracy with increasing field $h$ and $\Delta t$. Since in the discrete representation (10) the field action reduces to a rotation by an angle $h\Delta t$, then the accuracy decreases, if, with increasing $\Delta t$ and $h$, the angle increases to $\pi/2$.

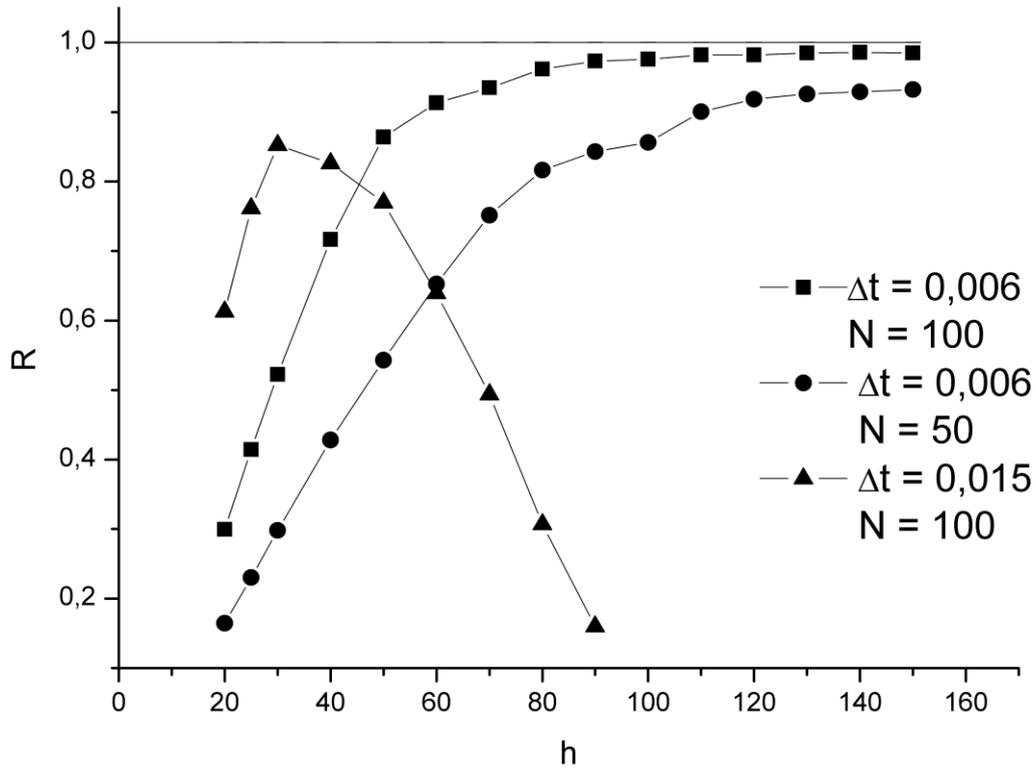

Figure 2 Dependences of the accuracy $R$ on the magnetic field $h$

The decrease in the accuracy on Fig.3 for small $\Delta t$ is due to the violation of adiabaticity. The accuracy decrease for large $\Delta t$ occurs as a result of the replacement of the continuous change of the Hamiltonian by a discrete one.

On the basis of numerical simulation, it is possible to select optimal parameters for the experimental implementation of the algorithm. According to Fig. 1 for $N = 10$, $\Delta t = 0.01$ and $h = 100$, the accuracy of the result is $R = 0.37$. The accuracy can be increased to $R = 0.45$ if we take $\Delta t = 0.0087$ and $h = 160$. An additional increase in the accuracy to $R = 0.48$ is achieved after symmetrization[16] of the formula (10).

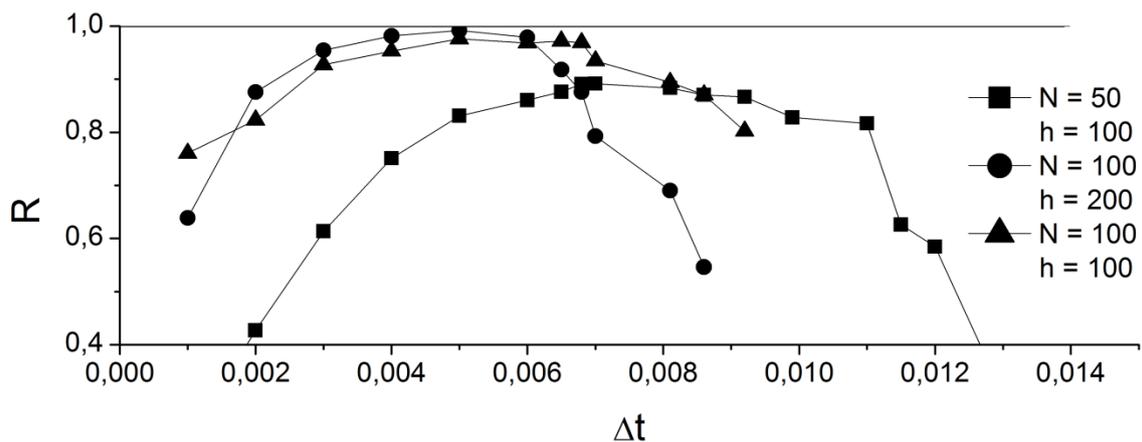

Fig.3 Dependences of the accuracy *R* on the duration of time step.

Thus, in this paper, we found sequences of selective rotation operators for creating one, two, and three spin interactions in a three-spin system $S = 1$. The possibilities of controlling the system of three qutrits are demonstrated by the example of factorizing the number 15 in the ternary number system by means of quantum annealing. The obtained formulas can be applied in the implementation of other quantum algorithms on qutrits.